# Why shot noise does not generally detect pairing in mesoscopic superconducting tunnel junctions

*Jiasen Niu,[1] Koen M. Bastiaans,[2] Jianfeng Ge,[1] Ruchi Tomar,[3] John Jesudasan,[3] Pratap Raychaudhuri,[3] Max Karrer,[4] Reinhold Kleiner,[4] Dieter Koelle,[4] Arnaud Barbier,[5] Eduard F.C. Driessen,[5] Yaroslav M. Blanter,[2] Milan P. Allan[1,6,7]\**

[1] *Leiden Institute of Physics, Leiden University, 2333 CA Leiden, The Netherlands*
[2] *Department of Quantum Nanoscience, Kavli Institute of Nanoscience, Delft University of Technology, 2628 CJ Delft, The Netherlands*
[3] *Department of Condensed Matter Physics and Materials Science, Tata Institute of Fundamental Research, Homi Bhabha Road, Colaba, Mumbai 400005, India*
[4] *Physikalisches Institut, Center for Quantum Science (CQ) and LISA+, Universität Tübingen, D-72076 Tübingen, Germany*
[5] *Institutde Radioastronomie Millim´etrique(IRAM),Domaine Universitairede Grenoble, 38400 Saint-Martin-d'H`eres, France*
[6] *Fakultät fürPhysik, Ludwig-Maximilians-Universität,Schellingstrasse4,München80799,Germany*
[6] *MunichCenter for Quantum Science and Technology(MCQST), München, Germany*
*\*milan.allan@lmu.de*

*The shot noise in tunneling experiments reflects the Poissonian nature of the tunneling process. The shot noise power is proportional to both the magnitude of the current and the effective charge of the carrier. Shot-noise spectroscopy thus enables – in principle – to determine the effective charge q of the charge carriers that tunnel. This can be used to detect electron pairing in superconductors: in the normal state, the noise corresponds to single electron tunneling ($q = 1e$), while in the paired state, the noise corresponds to $q = 2e$. Here, we use a newly developed amplifier to reveal that in typical mesoscopic superconducting junctions, the shot noise does not reflect the signatures of pairing and instead stays at a level corresponding to $q = 1e$. We show that transparency can control the shot noise and this $q = 1e$ is due to the large number of tunneling channels with each having very low transparency. Our results indicate that in typical mesoscopic superconducting junctions one should expect $q = 1e$ noise, and lead to design guidelines for junctions that allow the detection of electron pairing.*

For conventional superconductors, condensation and pairing occur concurrently when cooling below the superconducting critical temperature ($T_c$) [1]. However, for some unconventional [2,3] and disordered [4] superconductors, a state that contains phase incoherent, preformed pairs has been conjectured. Despite a wealth of tantalizing signatures, it is very difficult to distinguish preformed pairs from single electrons because (i), many experimental techniques to measure charge cannot differentiate between two single electrons and one pair, (ii), putative spectroscopic signatures like the pseudogap can have a variety of origins [5–7], and (iii), many experiments are dependent on models to determine whether they indicate paired or single electrons [2]. Yet, to test hypotheses of preformed pairs and bosonic liquids, direct and quantitative experimental information is needed in order to distinguish between a small fraction of fluctuating pairs and a liquid of bosonic particles.

In principle, shot noise spectroscopy is such an experimental technique. Because of the discrete nature of the charge carriers, the tunneling process is Poissonian, and thus the zero-temperature current noise power $S_I$ of a tunnel junction is proportional to the charge of each charge carrier $q$, and the absolute value of current $|I|$,

$$S_I = 2q|I|. \tag{1}$$

Therefore, shot noise can yield information on the effective charge in the system [8]. For a tunnel junction at finite temperature $T$, Eq. (1) is modified to $S_I = 2q|I|\coth\left(\frac{q|V|}{2k_BT}\right)$, where $V$ is the bias voltage across the tunnel junction, and $k_B$ is the Boltzmann constant. When the bias voltage is large compared to the temperature ($qV \gg k_BT$), it will reduce to Eq. (1). By measuring shot noise, the charge of the carrier can be directly determined. Shot noise has been used to detect the fractional effective charge in fractional quantum Hall effect [9,10], pairing in superconductors [11–14], and multiple Andreev reflection in superconducting tunnel junctions [14–16].

However, in mesoscopic superconducting junctions, with much larger areas where tunneling occurs (as opposed to atomic contacts in break junctions and scanning tunneling microscope (STM)) the interpretation of shot noise is not straightforward. For example, noise corresponding to $q = 1e$ has been measured in Nb/AlO$_x$/Nb tunnel junction, while enhanced noise ($q > 1e$) has been measured for NbN/MgO/NbN tunnel junction – in both cases, the samples are in the superconducting state [15]. Enhanced noise above $T_c$ was also measured in a La$_{2-x}$Sr$_x$CuO$_4$/La$_2$CuO$_4$/La$_{2-x}$Sr$_x$CuO$_4$ mesoscopic tunnel junction [17]. The interpretation of these surprising results is challenging: the noise is much larger than expected, and present even outside the superconducting gap. Taken together, it appears that the interpretation of shot noise experiments in mesoscopic superconducting junctions requires additional consideration. In the present work, we present careful measurements of mesoscopic superconducting junctions that yield $1e$-noise, and explain why these do not reveal signatures of pairing.

Before describing our experiments, we discuss the expectation for a typical voltage noise $S_V$ measurement in a tunnel junction (Fig. 1): it turns out to be very different from the standard V-shaped curve of current noise power $S_I$ seen in Eq. (1) and reported in current noise measurements in the literature [11,13,14]. The difference stems from the fact that in practice the voltage noise $S_V$ is measured instead of the current noise $S_I$. When biased with voltage $V$, the voltage noise is given by

$$S_V = S_I R_{diff}^2, \tag{2}$$

where $R_{diff}$ is the differential resistance of the sample. If the $I(V)$ curve is non-linear, like in a superconducting junction, the shape of the $S_V(V)$ curve will become complicated. To obtain a quantitative expectation, we can calculate $S_V$ for a superconductor-insulator-normal metal (SIN) tunnel junction using the Blonder-Tinkham-Klapwijk (BTK) formulas [14,18–20] (Fig. 1(d) ; for details see supplementary material). Our calculation shows that $S_V$ has a double peak structure inside the gap, the amplitude of which indicates the effective charge; below the superconducting gap, Andreev reflections lead to a doubling of the noise and hence the effective charge from $q^* = 1e$ to $q^* = 2e$ (Fig. 1(e)). Note that this model is for SIN junctions, which feature single Andreev reflections only. In contrast, superconductor-insulator-superconductor (SIS) junctions may give rise to multiple Andreev reflections [21], which further enhance the shot noise inside the superconducting gap at low energies [14–16]. Here we only consider single Andreev reflections.

Currently, there are two widely used state-of-the-art methods for measuring shot noise. The first method measures noise across a wide bandwidth at low frequencies using a room-temperature amplifier [17]. The key advantage of this approach is its broad bandwidth, enabling to distinguish frequency-independent shot noise from frequency-dependent 1/$f$ noise. The second method measures noise at high frequencies within a narrow bandwidth by employing an LCR resonator with an impedance matching circuit [13,14,22–24]. Its strength lies in its ability to mitigate the impact of 1/$f$ noise and random telegraph noise due to the higher operating frequency.

However, neither of these methods is suitable for the specific experimental challenges presented here. The first method has pronounced RC roll-off above 100 kHz [17], which makes it impossible to effectively extract shot noise from the total signal in our case. This is because the low-frequency portion of random telegraph noise exhibits a frequency-independent characteristic [25], similar to shot noise, and because the roll-off makes it difficult to measure small signals, considering that the values of capacitance and wire resistance may fluctuate with environmental changes. The second method – designed for system with resistances of the order of MΩ or more [13,14,22–24] – does not work for the low resistances (~Ω, due to the finite junction area) found in mesoscopic tunneling junctions as studied here [26]. The issue is that this leads to a quality factor that is too low to use the benefits of a LCR resonator.

We therefore built a new low-noise, high-frequency amplifier, designed to work in the frequency range of 100 kHz to 5 MHz, and to fulfill three main requirements: (i) a low-temperature environment to suppress the thermal noise from both the amplifier, (ii) a high-resolution amplifier to detect small signals, and (iii) low 1/$f$ noise. Details of the amplifier, including calibration methods and uncertainties of experimental parameters are given in the supplementary material.

We start our experiment with a planar Nb/Al-AlO$_x$/Nb (SIS) tunnel junction with a zero resistance $T_c$~8 K (Fig. 2). Tunneling spectroscopy shows the superconducting gap of the Bogoliubov density of states and the resulting non-linear $I(V)$ curve, which significantly influences our noise curves, as it is evident from Eq. **Error! Reference source not found.**. We deal with this challenge by first measuring the differential resistance $R_{diff}$ of the sample, and then calculating the expected shot noise for $q = 1e$ and $q = 2e$ according to

$$S_V = 2q|I|\coth(q|V|/(2k_BT))R_{diff}^2. \quad (3)$$

These expectations can then be compared with the experimental results, or to determine the effective charge from the experimentally measured noise at different bias voltages.

We start with noise data taken at higher temperatures, close to zero resistance $T_c$. The temperatures are measured by a calibrated thermometer and uncertainty of temperature measurements is around 0.01 K. Fig. 2(b) shows both the experimental data and the expected noise for $q = 1e$ and $q = 2e$, according to Eq. (3). It is obvious that the experimental data overlap very well with the $q = 1e$ theoretical line both outside and inside the gap. Next, we turn our attention to lower temperatures. Such measurements pose an additional complication due to the appearance of a supercurrent. As the differential resistance becomes zero, the voltage noise also reduces to zero according to Eq. (3) when there is supercurrent in the sample. To solve this problem, we apply an out-of-plane magnetic field of 0.1 T to suppress the supercurrent. The sample is in the center of the magnetic field and the homogeneity of the magnetic field is 0.1% over a 10 mm diameter spherical volume. With 0.1T, the supercurrent disappears but quasiparticles are induced as well; for details see supplementary material. From the theoretical calculation, shot noise will still be enhanced inside the gap even if there are more quasiparticles (normal tunneling process). However, we find that the noise still clearly corresponds $q = 1e$, as shown in Fig. 2(c). The effective charge stays at $q = 1e$ within the uncertainty (Fig. 2d).

An alternative method to avoid supercurrents is to use a SIN junction. We therefore measure noise at 8 K, 6 K, 4 K, and 2 K in a NbN/oxide/Ag SIN junction with $T_c \sim 15$ K(Fig. 3). For SIN junctions the zero bias $dV/dI$ is different to SIS junctions (Fig. 2a). For SIN junctions, when the temperature decreases below $T_c$ the gap will develop and the zero bias $dV/dI$ will increase. While for SIS junctions (Fig. 2), the $dV/dI$ will become zero due to the appearance of supercurrent. The experimental results at 2 K are shown in Fig. 3(b). Again, the measurements clearly indicate $q = 1e$ outside and inside the gap, even though the temperature $T = 2\ K$ is much lower than $T_c$. The noise characteristics at 8 K, 6 K, and 4 K are shown in Fig. S5; they also indicate that $q = 1e$.

In Fig. 4 we summarize the effective charge we measured inside the superconducting gap at different temperatures, different magnetic fields and different samples. Fig. 4 clearly shows that the effective charge inside the gap is 1$e$, even though all samples are superconducting. Details of samples and Fig. 4 are shown in supplementary material.

Our results raise the question why no 2$e$ charge transfer is observed in devices where the materials are clearly in their superconducting state. In theory [18,20,27], the appearance of Cooper pairs and Andreev reflections should yield doubled shot noise when the temperature is below $T_c$ in superconducting tunnel junctions, as previously observed in STM [13], nanowires [14], and break junctions [28].

One possibility to obtain $1e$-noise: tunneling through bound states in the insulator [29,30]. Such indirect tunneling leads to the possibility of normal tunneling inside the gap. However, it also leads to signatures in the tunneling spectra. As we do not observe these, we believe that this possibility is unlikely to be solely responsible for our results.

Instead, we argue here that this apparent inconsistency is due to the junction properties of typical mesoscopic setups, which are very different than those of STM [12,13,22] or nanowire [14]. In these systems, there is usually only one tunneling channel, and it is easy to deduce the transparency $\tau$ from $\tau = (G_0 R_j)^{-1}$, where $G_0$ is the conductance quantum and $R_j$ is the sample resistance. In contrast, our samples are mesoscopic tunnel junctions, made by cleanroom fabrication methods, have larger junction areas, e.g. 25 $\mu m^2$ for our SIS device. Because of this large junction area, the number $N$ of tunneling channels in our sample is large and it is a priori not possible to separately deduce the number of channels and the transparency, $\tau = (N G_0 R_j)^{-1}$.

It is important to point out that mesoscopic junctions generally have large numbers of channels $N$ with much smaller transparencies compared to point-like junctions with the same resistance. And indeed, such a situation can lead to vastly different shot noise, as we show by simulations (Fig. 5). The key point is that the shot noise is controlled by the transparency $\tau$. While a typical single channel junction with typical resistances will show $2e$ noise, a mesoscopic junction with typical parameters (but without pinholes) will show $1e$ noise, because of the vastly different transparencies. This can be illustrated with an example: a 30 Ω mesoscopic junction with an area of 0.82 $\mu m^2$ might have $N \sim 10^7$ channels, meaning that the transparency is roughly $10^{-5}$, if homogenous channel parameters are assumed. It is these small transparencies that explain our $1e$ noise. At finite temperatures Andreev reflections and normal tunneling process will happen at the same time inside the gap and the transparency will control the contribution from each of these two processes to the total noise. The contribution from normal process is proportional to $\tau$ while the contribution from Andreev reflections is proportional to $\tau^2$. Thus, when the transparency is low, the normal process will dominate the shot noise signal, giving $q = 1e$; only when the transparency is high, one can observe the noise enhancement. We generalize this further in Fig. 5(b) where we show simulations of the effective charge as a function of $N$ or $\tau$ for different temperatures. We use the parameters corresponding to our SIN sample. This simulation confirms that for an increasing number of channels (decreasing transparency), the effective charge decreases, and provides guidelines for obtaining $2e$ shot noise, as outlined below.

To determine where our experimental system is within this parameter range and test our hypothesis, we estimate the possible number of channels and their transparencies in our samples. The number of channels can be estimated by $N = A * \frac{k_F^2}{\pi}$, where $A$ is the area of the junction and $k_F$ is the Fermi wave vector [31]. $k_F$ can be estimated from $k_F = (3\pi^2 n)^{1/3}$, where $n$ is the charge number density in the bulk. The charge densities of Nb [32] and NbN [26] are estimated from the literature. In Table 1, we present our estimates for the numbers of channels and transparencies of our samples. When comparing these values with our model (Fig. 5(b)), we find that indeed our samples should yield $q = 1e$ values.

Thus far, we used BTK formulas to calculate the effective charge in SIN junctions. For SIS junctions, the situation is similar. We use a quasi-particle tunneling model [33] to calculate the contribution from $q = 1e$ channels and find that $q = 1e$ channels already dominates the tunneling processes due to the small $\tau$ in our sample (see supplementary materials). We also note that there exist multiple Andreev reflections in SIS junctions, which will lead to further enhanced shot noise. However, multiple Andreev reflections processes will be suppressed compared to Andreev reflections at low transparencies, as their probabilities scale with $\tau^n$.

We can now use our model to estimate $\tau$ and compare its prediction to different results reported in literature. We find that $q > 1e$ was reported in all systems with where we expect large $\tau$ ($\tau > \sim 10^{-4}$) [12–15,34] and $q = 1e$ was reported in most systems with small $\tau$ [15,35], in agreement with our model (details see Fig. S9). The exception stems from an experiment on a cuprate junction, where $q > 1e$ was reported for a very small value of $\tau$ [17]. It is an open question how to interpret this work; here we note that theory for noise of preformed pairs is still developing, and that the charge transfer layers can yield extra noise [36]. We also performed experiments on a $YBa_2Cu_3O_7$ high-$T_c$ superconductor junction [37], however, the noise spectrum in these samples is dominated by 1/$f$ noise (details see Fig. S8).

In summary, we measured shot noise in SIS (Nb/Al-AlO$_x$/Nb) and SIN (NbN/oxide/Ag) superconducting tunnel junctions in a high-frequency bandwidth (1.1-1.5 MHz) using a new, custom-built high-resolution noise measurement system. We find that the measured effective charge equals 1$e$ both outside and inside the energy gap at temperatures below $T_c$ for both types of junctions. This could be considered counterintuitive as in principle the noise in the superconducting state should correspond to $q = 2e$ because of Andreev reflections. We interpret our findings by proposing the presence of a large number of very small transparency channels in our sample, which obscure the pairing effect in noise measurements. BTK simulations quantitatively agree with this picture. We further argue that this is in fact a common situation for mesoscopic junctions.

Our findings therefore indicate that to measure electron pairs without superconductivity in mesoscopic junctions – as opposed to single-channel STM junctions – one needs to carefully engineer a junction with few channels with very high transparency. Junctions with very thin insulator layers and clean interfaces are good candidates to achieve such a scenario. Further, pinholes in junctions, which are usually to be avoided, may give extra high transparency ($q = 2e$) channels.

**ACKNOWLEDGMENTS**

We acknowledge C. W. J. Beenakker for valuable discussions. This work was supported by the European Research Council (ERC StG SpinMelt and ERC CoG PairNoise). K.M.B. was supported by the Dutch Research Council (Veni Grant No. VI. Veni. 212.019). Y. M. B. was supported by European Space Agency (ESA) under ESA CTP Contract No.4000130346/20/NL/BW/os. R.T., J.J., and P.R. were supported by the Department of Atomic Energy, Government of India (Grant No. 12-R&D-TFR5.10-0100).

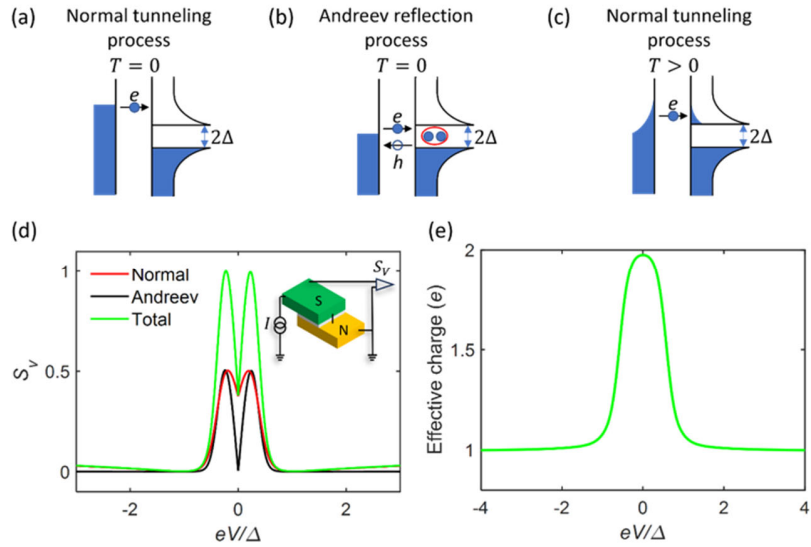

Fig. 1: Tunneling process and shot noise in a NIS tunnel junction. (a) Illustration of single electron tunneling at biases larger than the gap. Black lines indicate the density of states of the superconductor and the normal metal; filled states are shaded in blue. (b) Andreev reflection process when the bias voltage is smaller than the gap, leading to 2*e*-shot noise. (c) At finite temperatures, normal tunneling processes occur inside the gap. (d) Normalized voltage noise power calculated for different process ($\Delta = 3.8$ meV, $T = 4$ K, and transparency $\tau = 0.01$). Red line: noise curve assuming only normal tunneling processes. The finite value at zero bias is thermal noise. Black line: shot noise from Andreev reflection processes. Green line: total noise, which is what is measured in experiments. Inset: schematic of the electrical circuit. (e) Effective charge calculated from the total noise.

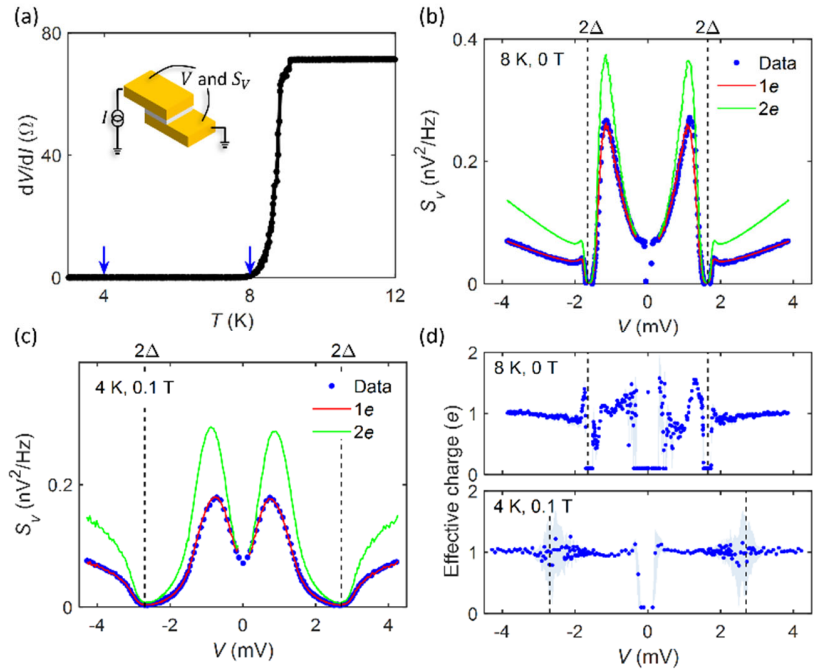

Fig. 2: Shot noise measurements on SIS junctions. (a) Zero bias $dV/dI$ from Nb/Al-AlO$_x$/Nb sample C1. The blue arrows indicate the temperatures of the noise measurements shown in (b), (c); the inset shows the schematic of the sample. (b) Measured voltage noise data (blue dots) and calculations of the shot noise with $q = 1e$ (red line) and $q = 2e$ (green line), at 8 K and zero magnetic field. Black dashed lines indicate the superconducting gap. (c) Noise measurements at 4 K and 0.1 T. (d) Effective charge at different bias voltages. The shaded areas indicate the uncertainty of the data.

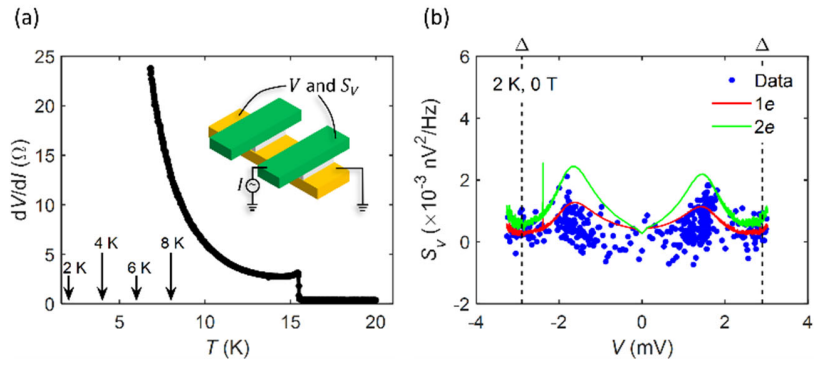

Fig. 3: Noise measurement from a NbN/oxide/Ag (SIN) junction N1. (a) Zero bias $dV/dI$ measured with constant AC current of 0.25 mA. Inset: top view of the sample. The yellow color indicates NbN; the green colors indicates Ag. Arrows indicate the temperatures at which noise spectra have been measured. Noise data at 2 K without magnetic field are shown in (b); data of other temperatures are shown in supplementary materials.

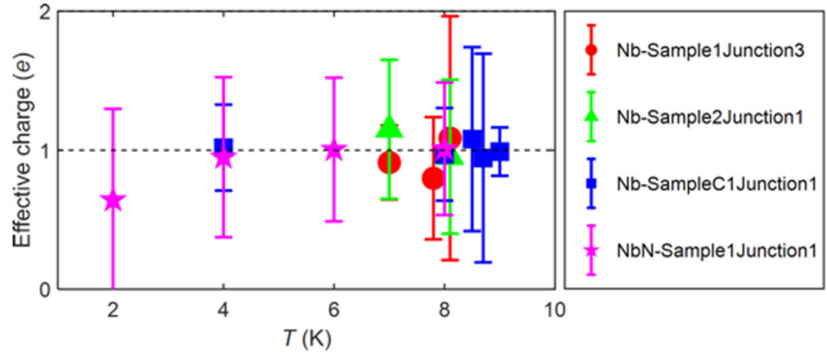

Fig. 4: Summary of the averaged effective charge inside the superconducting gap for different samples as a function of temperature. The zero resistance $T_c$ is ~8 K for Nb based samples and 15.5 K for NbN based samples. The effective charge of the Nb/Al-AlO$_x$/Nb sample at 4 K and 7 K is measured at 0.1 T.

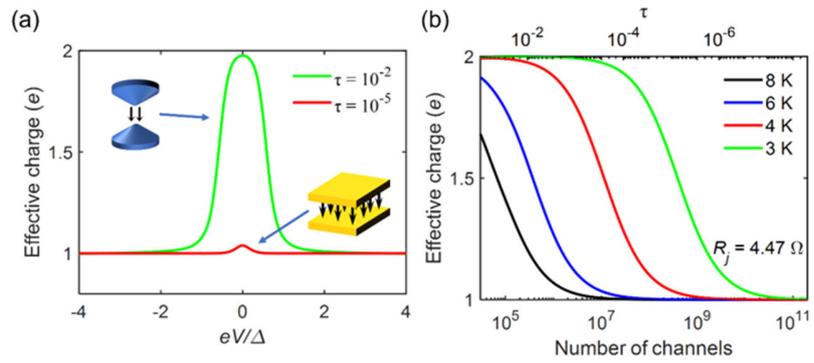

Fig. 5: Obscuring of 2*e* noise in mesoscopic device geometries. (a) Effective charge calculated using BTK theory for transparencies of $10^{-2}$ (typical for STM geometries; green line) and $10^{-5}$ (typical for mesoscopic devices; red line). (b) Effective charge inside the gap as a function of the number of channels (bottom axis) and transparency (top axis) for a sample with $4.47\ \Omega$ normal state resistance for different temperatures.

**Table 1 Transparency estimation**

| Sample | Nb/AlOx/Nb | | NbN/oxide/Ag |
|---|---|---|---|
| | sample 1,2 | sample C1 | sample N1 |
| charge density | $5.56\times10^{28} m^{-3}$ | | $1.60\times10^{29} m^{-3}$ |
| R | 31.6 Ω | 63.1 Ω | 4.47 Ω |
| Area | 0.82 $\mu m^2$ | 25 $\mu m^2$ | 0.09 mm$^2$ |
| $\tau$ | $1.12\times10^{-5}$ | $1.85\times10^{-7}$ | $3.70\times10^{-10}$ |
| N | $3.64\times10^{7}$ | $1.11\times10^{9}$ | $7.80\times10^{12}$ |

[1] M. Tinkham, *Introduction to Superconductivity*, Courier Corporation (2004).

[2] B. Keimer, S. A. Kivelson, M. R. Norman, S. Uchida, and J. Zaanen, *From Quantum Matter to High-Temperature Superconductivity in Copper Oxides*, Nature **518**, 179 (2015).

[3] V. J. Emery and S. A. Kivelson, *Importance of Phase Fluctuations in Superconductors with Small Superfluid Density*, Nature **374**, 434 (1995).

[4] B. Sacépé, M. Feigel'man, and T. M. Klapwijk, *Quantum Breakdown of Superconductivity in Low-Dimensional Materials*, Nat Phys **16**, 734 (2020).

[5] T. Timusk and B. Statt, *The Pseudogap in High-Temperature Superconductors: An Experimental Survey*, Reports on Progress in Physics **62**, 61 (1999).

[6] M. V. Sadovskii, *Pseudogap in High-Temperature Superconductors*, Physics-Uspekhi **44**, 515 (2001).

[7] A. A. Kordyuk, *Pseudogap from ARPES Experiment: Three Gaps in Cuprates and Topological Superconductivity (Review Article)*, Low Temperature Physics **41**, 319 (2015).

[8] Y. M. Blanter and M. Büttiker, *Shot Noise in Mesoscopic Conductors*, Phys Rep **336**, 1 (2000).

[9] R. De-Picciotto, M. Reznikov, M. Heiblum, V. Umansky, G. Bunin, and D. Mahalu, *Direct Observation of a Fractional Charge*, Nature **389**, 162 (1997).

[10] L. Saminadayar, D. C. Glattli, Y. Jin, and B. Etienne, *Observation of the e/3 Fractionally Charged Laughlin Quasiparticle*, Phys Rev Lett **79**, 2526 (1997).

[11] X. Jehl, M. Sanquer, R. Calemczuk, and D. Mallly, *Detection of Doubled Shot Noise in Short Normal-Metal/ Superconductor Junctions*, Nature **405**, 50 (2000).

[12] K. M. Bastiaans, D. Cho, D. Chatzopoulos, M. Leeuwenhoek, C. Koks, and M. P. Allan, *Imaging Doubled Shot Noise in a Josephson Scanning Tunneling Microscope*, Phys Rev B **100**, 104506 (2019).

[13] K. M. Bastiaans et al., *Direct Evidence for Cooper Pairing without a Spectral Gap in a Disordered Superconductor above Tc*, Science **374**, (2021).

[14] Y. Ronen, Y. Cohen, J. H. Kang, A. Haim, M. T. Rieder, M. Heiblum, D. Mahalu, and H. Shtrikman, *Charge of a Quasiparticle in a Superconductor*, Proceedings of the National Academy of Sciences **113**, 1743 (2016).

[15] P. Dieleman, H. G. Bukkems, T. M. Klapwijk, M. Schicke, and K. H. Gundlach, *Observation of Andreev Reflection Enhanced Shot Noise*, Phys Rev Lett **79**, 3486 (1997).

[16] J. C. Cuevas, A. Martín-Rodero, and A. L. Yeyati, *Shot Noise and Coherent Multiple Charge Transfer in Superconducting Quantum Point Contacts*, Phys Rev Lett **82**, 4086 (1999).

[17] P. Zhou, L. Chen, Y. Liu, I. Sochnikov, A. T. Bollinger, M. G. Han, Y. Zhu, X. He, I. Božović, and D. Natelson, *Electron Pairing in the Pseudogap State Revealed by Shot Noise in Copper Oxide Junctions*, Nature **572**, 493 (2019).

[18] G. E. Blonder, M. Tinkham, and T. M. Klapwijk, *Transition from Metallic to Tunneling Regimes in Superconducting Microconstrictions: Excess Current, Charge Imbalance, and Supercurrent Conversion*, Phys Rev B **25**, 4515 (1982).

[19] M. P. Anantram and S. Datta, *Current Fluctuations in Mesoscopic Systems with Andreev Scattering*, Phys Rev B **53**, 16390 (1996).

[20] M. Octavio, M. Tinkham, G. E. Blonder, and T. M. Klapwijk, *Subharmonic Energy-Gap Structure in Superconducting Constrictions*, Phys Rev B **27**, (1983).

[21] T. M. Klapwijk, G. E. Blonder, and M. Tinkham, *Explanation of Subharmonic Energy Gap Structure in Superconducting Contacts*, Physica B+C **109–110**, 1657 (1982).

[22] K. M. Bastiaans, T. Benschop, D. Chatzopoulos, D. Cho, Q. Dong, Y. Jin, and M. P. Allan, *Amplifier for Scanning Tunneling Microscopy at MHz Frequencies*, Review of Scientific Instruments **89**, 93709 (2018).

[23] A. Bid, N. Ofek, M. Heiblum, V. Umansky, and D. Mahalu, *Shot Noise and Charge at the 2/3 Composite Ffractional Qquantum Hall State*, Phys Rev Lett **103**, 236802 (2009).

[24] U. Kemiktarak, T. Ndukum, K. C. Schwab, and K. L. Ekinci, *Radio-Frequency Scanning Tunnelling Microscopy*, Nature **450**, 85 (2007).

[25] L. Liu, L. Xiang, H. Guo, J. Wei, D. L. Li, Z. H. Yuan, J. F. Feng, X. F. Han, and J. M. D. Coey, *Low Frequency Noise Peak near Magnon Emission Energy in Magnetic Tunnel Junctions*, AIP Adv **4**, 12 (2014).

[26] S. P. Chockalingam, M. Chand, A. Kamlapure, J. Jesudasan, A. Mishra, V. Tripathi, and P. Raychaudhuri, *Tunneling Studies in a Homogeneously Disordered s -Wave Superconductor: NbN*, Phys Rev B **79**, 094509 (2009).

[27] C. W. J. Beenakker, *Random-Matrix Theory of Quantum Transport*, Rev Mod Phys **69**, 731 (1997).

[28] R. Cron, M. F. Goffman, D. Esteve, and C. Urbina, *Multiple-Charge-Quanta Shot Noise in Superconducting Atomic Contacts*, Phys Rev Lett **86**, 4104 (2001).

[29] A. L. Fauchè, G. B. Lesovik, and G. Blatter, *Finite-Voltage Shot Noise in Normal-Metal-Superconductor Junctions*, Phys Rev B **58**, 17 (1998).

[30] J. Nilsson, A. R. Akhmerov, and C. W. J. Beenakker, *Splitting of a Cooper Pair by a Pair of Majorana Bound States*, Phys Rev Lett **101**, 120403 (2008).

[31] D. Esteve, J.-M. Raimond, and J. Dalibard, *Quantum Entanglement and Information Processing: Lecture Notes of the Les Houches Summer School 2003*, Elsevier 509 (2004).

[32] N. Ashcroft and N. Mermin, *Solid State Physics*, Cengage Learning (2022).

[33]  R. C. Dynes, V. Narayanamurti, and J. P. Garno, *Direct Measurement of Quasiparticle-Lifetime Broadening in a Strong-Coupled Superconductor*, Phys Rev Lett **41**, 1509 (1978).

[34]  J. F. Ge, K. M. Bastiaans, D. Chatzopoulos, D. Cho, W. O. Tromp, T. Benschop, J. Niu, G. Gu, and M. P. Allan, *Single-Electron Charge Transfer into Putative Majorana and Trivial Modes in Individual Vortices*, Nat Commun **14**, 3341 (2023).

[35]  F. Massee, Q. Dong, A. Cavanna, Y. Jin, and M. Aprili, *Atomic Scale Shot-Noise Using Cryogenic MHz Circuitry*, Review of Scientific Instruments **89**, 9 (2018).

[36]  K. M. Bastiaans, D. Cho, T. Benschop, I. Battisti, Y. Huang, M. S. Golden, Q. Dong, Y. Jin, J. Zaanen, and M. P. Allan, *Charge Trapping and Super-Poissonian Noise Centres in a Cuprate Superconductor*, Nat Phys **14**, 1183 (2018).

[37]  B. Müller, M. Karrer, F. Limberger, M. Becker, B. Schröppel, C. J. Burkhardt, R. Kleiner, E. Goldobin, and D. Koelle, *Josephson Junctions and SQUIDs Created by Focused Helium-Ion-Beam Irradiation of YBa2Cu3O7*, Phys Rev Appl **11**, 044082 (2019).

## Supplemental material

**Table of Contents**

**I: Calculation of shot noise in SIN junctions using the BTK formulas**

Following previous work [1–3], we calculate the *I-V* curve and spectral density of noise power *S* in a superconductor-insulator-normal metal (SIN) tunnel junction by using the Blonder-Tinkham-Klapwijk (BTK) formulas. According to [2], the *I-V* curve of normal-superconducting system can be calculated by:

$$I_{NS} = 2N(0)ev_F \mathcal{A} \int_{-\infty}^{\infty} [f_0(E - eV) - f_0(E)][1 + A(E) - B(E)]dE, \quad (S1)$$

where $N(0)$ is the one-spin density of states at the Fermi energy $E_F$, $E$ is the absolute energy minus $E_F$, $e$ is the electron charge, $v_F$ is the Fermi velocity, $\mathcal{A}$ is the effective neck cross-sectional area, $f_0(E)$ is the Fermi-Dirac distribution function, $A(E)$ is the coefficient of Andreev reflection, and $B(E)$ is the probability of ordinary reflection. $A(E)$ and $B(E)$ can be calculated by:

|  | $A(E)$ | $B(E)$ |
|---|---|---|
| $E < \Delta$ | $\dfrac{\Delta^2}{E^2 + (\Delta^2 - E^2)(1 + 2Z^2)^2}$ | $1 - A$ |
| $E > \Delta$ | $\dfrac{u_0^2 v_0^2}{\gamma^2}$ | $\dfrac{(u_0^2 - v_0^2)^2 Z^2(1 + Z^2)}{\gamma^2}$ |

Here, $\gamma^2 = [u_0^2 + Z^2(u_0^2 - v_0^2)]^2$, and $\Delta$ is the energy gap. $u_0$ and $v_0$ can be calculated by $u_0^2 = 1 - v_0^2 = \frac{1}{2}\{1 + [(E^2 - \Delta^2)/E^2]^{1/2}\}$. $Z$ is the dimensionless barrier strength and can be calculated by $Z = \sqrt{1/\tau - 1}$, where $\tau$ is the transparency of the tunnel junction at normal state.

The current shot noise power $S_I$ can be calculated by [1,3]:

$$S_I = 4N(0)e^2 v_F \mathcal{A} \int_{-\infty}^{\infty} [f_0(E-ev) - f_0(E)] \begin{bmatrix} A(E)[1+A(E)] + \\ B(E)[1-B(E)] + 2A(E)B(E) \end{bmatrix} dE. \quad (S2)$$

Finally, the total voltage noise power $S_V$, the green line in Fig. 1(d) of the main text, is:

$$S_V = S_I R_{diff}^2, \quad (S3)$$

and the differential resistance $R_{diff} = dV/dI$ can be calculated from Eq. (S1).

If there is no Andreev reflection and only normal tunneling process, $A(E) = 0$, and the expression for shot noise becomes:

$$S_{V-1e} = \frac{4e^2}{h} R_{diff}^2 \int_{-\infty}^{\infty} [f_0(E-ev) - f_0(E)][B(E)[1-B(E)]]dE, \quad (S4)$$

which is the red line in Fig. 1(d).

If there is Andreev reflection, this will give rise to an extra contribution to the shot noise:

$$S_{V-extra} = \frac{4e^2}{h} R_{diff}^2 \int_{-\infty}^{\infty} [f_0(E-ev) - f_0(E)] \begin{bmatrix} B(E)[1-B(E)] + \\ 2A(E)B(E) \end{bmatrix} dE, \quad (S5)$$

which is the black line in Fig. 1(d).

Finally, the Fano factor $F$ can be calculated by:

$$F = \frac{S_I}{2eI}. \quad (S6)$$

In our model, we only consider SIN junctions without multiple Andreev reflection. In this case, the effective charge $q$ is equal to the Fano factor:

$$q = \frac{S_I}{2eI}. \quad (S7)$$

Combining Eqs (S1), (S2), and (S7) gives rise to the plot shown in Fig. 1(e) of the main text.

**II: Measurement methods**

### A: Low temperature noise amplifier circuit

In this work, we built a new low-noise, high-frequency amplifier, designed to work in the frequency range from 100 kHz to 5 MHz, and to fulfill three main requirements: (i) a low-temperature environment to suppress the thermal noise from both the sample and the amplifier, (ii) a high-resolution amplifier to detect small signals, and (iii) low 1/*f* noise. Fig. S1(a) shows the circuit including a commercial ATF-35143 transistor, and Fig. S1(b) shows a photograph of the amplifier. A commercial voltage amplifier (FEMTO HAS-X-1-40) and high-frequency SMA connectors are used at room temperature. After the room-temperature amplifier, a spectrum analyzer (Zurich MFLI) is used to digitize the voltage noise. The measurement system is built in a TeslatronPT from Oxford Instruments, in which the temperature can change from 1.5 K to 300 K. Detailed information on the low temperature amplifier design is discussed in next section II-B.

Fig. S1(c) shows an experimental spectrum for thermal noise of a 1 kΩ resistor at 1.59 K. The spectral voltage noise in Fig. S1(c) is $\frac{S_{measured}}{Gain_{RT}}$, where $S_{measured}$ is what we measure from the spectrum analyzer and $Gain_{RT} = 10000$ is the power gain of the room temperature amplifier (FEMTO HAS-X-1-40). It includes both thermal noise of the 1 kΩ ($4k_B TR$), background noise from amplifiers and spectrum analyzer ($BG$) and HEMT gain ($Gain_{HEMT}$): $\frac{S_{measured}}{Gain_{RT}} = Gain_{HEMT} * (4k_B TR + BG)$. The spectrum is very clean, with nearly no extra noise and the RC roll-off is not strong below 3 MHz. We can measure shot noise in the MHz region and estimate $1/f$ noise from the hundred kHz region. In this paper, the noise is measured from 1.1 MHz to 1.5 MHz, as indicated by the black dashed lines in Fig. S1(c).

## B: Amplifier calibration

The gain of the noise amplifier, $Gain_{HEMT}$, is calibrated by looking at the thermal noise. A resistor $R = 1$ kΩ is put on the sample holder as noise source. The resistance we used is thin metal film resistor which changes little with temperature and its resistance at 2 K was measured for double check by a four probe measurement before noise amplifier calibration. The temperature is measured by a calibrated thermometer near the sample and exchange gas is used to ensure that the temperature is homogenous. The uncertainty of temperature is about 0.01 K.

To calibrate the power gain, we first measure noise power $S_{measured}$ by a spectrum analyzer at different sample temperatures and then the power gain of HEMT, $Gain_{HEMT}$, can be calculated by: $\frac{S_{measured}}{Gain_{RT}} = (4Gain_{HEMT} k_B R) * T + Gain_{HEMT} * BG$. Then we can measure noise at different temperature and fit $\frac{S_{measured}}{Gain_{RT}}$ vs $T$. The slope of $\frac{S_{measured}}{Gain_{RT}}$ vs $T$ is $4Gain_{HEMT} k_B R$. $k_B$ and $R$ are known and then $Gain_{HEMT} = 1.25^2 \approx 1.56$ can be calculated. Then,

$$S_V = S_{measured}/Gain_{RT}/Gain_{HEMT} = 4k_B TR + BG, \tag{S8}$$

can be used to calculate the effective BG. $S_V$ vs $T$ is plotted in Fig.S1(d) and the intercept of the dashed line is effective $BG = 0.28$ nV$^2$/Hz. The uncertainty of $Gain_{HEMT}$ and BG are 0.01 and 0.006nV^2/Hz.

Our amplifier power gain, $Gain_{HEMT}$, is an effective value. The resistance and capacitance in our system are taken into account in our calculation and other parts are included in the effective power gain. To exclude the mistake from measurements, our noise measurement system is specially designed.

As shown in Fig. 1(a), $C_w$ is cable capacitance of the cable between low temperature and room temperature, and $R_{LT}$ are low temperature resistors directly connected to the sample holder. $R_{LT}$ is designed to be significantly larger than the sample differential resistance (for both low frequency and MHz frequency), allowing it to be safely disregarded in calculations. This design also helps prevent any influence from $C_w$. $C_E$ is an effective capacitance. It includes cable capacitance between sample and amplifier and the capacitance from HEMT gate to HEMT source. $C_c$ is a capacitor on the amplifier used to block DC current to HEMT. $R_{Amp}$ is the amplifier input impedance which is much larger than the sample resistance. Impedance of $C_E$ and $R_{Amp}$ are taken into account in our calculation. And many filters are used to avoid the extra noise from room temperature equipment.

## C: Differential resistance measurement

Our differential resistance $R_{diff} = dV/dI$ is measured with an AC technique. We combine a signal generator with low temperature resistor to apply a DC bias current $I$ and AC current $dI = 0.3$μA at 113 Hz on the sample at the same time. The signal is then measured by a low frequency preamplifier (NF LI-75) and lock-in (Stanford Research SR830). We measure $R_{diff}$ at the same magnetic field, same temperature and same speed as the noise measurement. This means that for each noise curve a $R_{diff}$ curve is also measured with the same parameters.

We also measured $R_{diff}$ at 1.1 MHz, same frequency with our noise measurement. We find that $R_{diff}$ at different frequencies can overlap with each other. This is because that we measure noise at 1.1-1.5 MHz instead of GHz. At this frequency, the resistance of the sample is similar to the one at low frequency. So we can use $R_{diff}$ measured at 113 Hz in our noise calculation. Similar frequency independent $R_{diff}$ at MHz was also reported in previous work [4,5]. Please notice that the quantity we discussed above is the differential resistance $R_{diff}$, not the total impedance of sample and cables. The total impedance is different for different frequencies because of the capacitance. To solve this possible problem, our noise measurement system is specially designed to suppress the influence from the capacitance.

**III: Nonlinear *I-V* and d*I*/d*V* curves**

Fig. S2 shows the *I-V* and d*I*/d*V* curves of superconductor-insulator-superconductor (SIS) and SIN junctions discussed in the main text. Tunneling spectroscopies show a good superconducting gap. We divide the current noise power by $R_{diff}^2$ according to Eq. (3) to calculate the voltage noise power. The nonlinear *I-V* curves lead to the double peak shape of the measured voltage power graphs structure as well as of the 1*e* and 2*e* theory lines in Figs. 1, 3, and 4 of the main text.

**IV Use of a magnetic field to suppress the supercurrent**

One kind of sample we used is a Nb/Al-AlO$_x$/Nb tunnel junction. In this sample a supercurrent will appear when the temperature is much lower than $T_c$. In that case, it is not possible to measure the shot noise because, as the sample resistance drops to zero, the voltage shot noise also reduce to zero, according to Eq. (3). To solve this problem, we apply a small magnetic field to suppress the supercurrent when we measure shot noise far below $T_c$. Fig. S3(a) shows the *I-V* curves for "SampleC1" at 4 K. Without magnetic field (red curve), the voltage will directly drop to zero inside the gap. With a magnetic field of 0.1 T (green curve) and 0.2 T (red curve), the supercurrent disappears, allowing the shot noise to be measured at 4 K. And quasiparticles are induced at the same time according to *I-V* curves. Noise measurement results for 4 K and 0.1 T are shown in Fig. 3(c) of the main text. Fig. S3(b) shows the results for 4 K and 0.2 T. As for 4 K and 0.1 T (main text Fig. 3(c)), the shot noise for 4 K and 0.2 T also clearly shows $q = 1e$ behavior with no magnetic field dependence.

**V: Shot noise near $T_c$**

In the main text, $q = 1e$ is measured below $T_c$. For comparison, shot noise near $T_c$ is also measured and shown in Fig. S4. Fig. S4(a) shows the *R-T* curve for a SIS junction ("SampleC1") with arrows indicating the temperatures at which the noise was measured (8.5 K, 8.7 K, and 9K). Inset shows the tunneling spectroscopy for different temperatures. At 8.5 K, spectroscopy still shows gap behavior. At 9 K the gap nearly disappears. Figs. S4(b-d) show shot noise results at different temperatures. For all these temperatures near $T_c$, the experimental data overlap with theoretical $q = 1e$ lines, similar to what happens well below $T_c$.

**VI: 1*e*-noise at different temperatures in a SIN sample**

For SIN samples, shot noise was measured at different temperatures. The results for the lowest temperature (2 K) are shown in Fig. 4 of the main text. Other results for SIN samples are shown here in Fig. S5. Similar to Fig. S4, Fig. S5(a) shows the *R-T* curve and tunneling spectroscopies. Arrows indicate the temperatures at which the noise was measured (8 K, 6 K, 4 K, and 2 K). Fig. S5(b-d) show the noise results at different temperatures, all of which agree with $q = 1e$.

**VII: Calculation of the average effective charge**

To calculate averaged effective charges, we make use of the experimentally measured voltage noise power, $S_V$, as a function of bias voltage. Typical data are shown in blue dots in Fig. S6(a) and (b). Then effective charge as a function of voltage can be calculated by:

$$S_I = 2qIR_{diff}^2 \coth\left(\frac{qV}{2k_BT}\right). \tag{S9}$$

Per sample and temperature, we then average all the effective charges over the voltage domains inside the gap shown in green in Fig. S6(c,d). The results are plotted in Fig. 5 of the main text. It can be seen that the $S_V$ for $V = 0$ and $V = \Delta$ is very small. In these voltage domains (indicated by red circles in Fig. S6(c,d)) the effective charge is difficult to measure, and therefore not used in the calculation.

**VIII: Noise calculation for SIS junctions**

In the main text our model focuses on the SIN junctions and can explain our results in SIN junctions well. Our model can calculate the results in SIN junction fully quantitatively. The contribution from Andreev reflection and normal tunneling process with different transparency are well calculated. We believe that the calculations for SIS junctions should not differ significantly from SIN junctions, since we can also separate the tunneling process into normal tunneling process ($q = 1e$), Andreev reflection process ($q = 2e$) and n-th order Andreev reflection process ($q = ne$) and calculate contribution from each process separately [16, 28]. Thus, we expect the transparency to play a qualitatively similar role in controlling shot noise in SIS junctions as in SIN junctions.

Unfortunately, we have not yet been able to find an good method for quantitatively simulating the total shot noise from all the tunneling processes in SIS junctions that is similar to what we did for SIN junctions. Sill, we can use a way to understand our results: instead of calculating the total noise, we can calculate the contribution of $q = 1e$ noise from the normal tunneling process.

We can use the Bardeen tunneling theory to calculate the normal tunneling process [6]:

$$I = I_n \int_{-\infty}^{+\infty} D_{1S}(\varepsilon) D_{2S}(\varepsilon + eV)[F_{FD}(\varepsilon) - F_{FD}(\varepsilon + eV)]d\varepsilon, \tag{S10}$$

where $I$ is the tunneling current, $I_n$ is normal state tunneling current, $D_{1S}(\varepsilon)$ and $D_{2S}(\varepsilon + eV)$ are the Dynes functions of the left and right superconducting materials, $F_{FD}(\varepsilon)$ and $F_{FD}(\varepsilon + eV)$ are the Fermi-Dirac distribution functions of the two sides. From this current, the noise power from normal tunneling processes in an SIS junction can be calculated by using Eq. (S9). Fig. S7 shows that the theoretical calculation and experimental data of Nb/Al-AlO$_x$/Nb at 8 K without a magnetic field overlap well. This means that the tunneling process in our SIS samples is dominated by $q = 1e$ with small transparency and so $q = 1e$ noise will be measured in our SIS samples.

We also note that there are multiple Andreev reflections (MAR) in SIS junctions. MAR will lead to enhanced shot noise even larger than Andreev reflection. But at the same time, MAR noise will be suppressed more by transparency effects, as compared to Andreev reflection because $q = 2e$ noise is suppressed by $\tau^2$ while $q = ne$ noise will be suppressed by $\tau^n$ with $n > 2$. Thus MAR may lead to larger noise but it is even more difficult to be detected with small $\tau$.

**IX: Shot noise in the high-$T_c$ superconductor YBa$_2$Cu$_3$O$_7$**

We also measured noise in high-$T_c$ superconductor YBa$_2$Cu$_3$O$_7$. The sample is a Josephson junction fabricated by a focused helium ion beam. The inset of Fig. S8 shows the sample structure. The fabrication method is reported in previous work [7]. Unfortunately, 1/$f$ noise still dominates the noise spectrum at MHz frequencies, preventing us from accurately measuring the shot noise. Fig. S8 shows a typical noise spectrum measured in YBa$_2$Cu$_3$O$_7$ at 50 μA bias current.

**X: Distribution of transmission probabilities in a tunnel junction in series with a diffusive conductor**

Our sample is actually made up of a tunnel junction and a diffusive conductor in series. Even though four probe measurements have been used, there is still some contribution from the conductors in series to the measurements. In the calculation, we assumed that all transmission eigenvalues are very small and equal to each other. However, the presence of a diffusive conductor results in a distribution of transmission eigenvalues. In this Supplementary Section. We investigate this distribution and conclude that it does not change our results.

To estimate the contribution from diffusive conductors, we need to calculate the average transmission $\langle T \rangle$ for the system. This problem is actually solved in the book by Nazarov and Blanter [8] (see Eq(2.115) of the book), but the obtained result is not explicit. We produce below explicit results for $G_D \gg G_T$, where $G_D$ and $G_T$ are total conductance of the diffusive conductor and the tunnel junction, respectively.

For the distribution function $\rho(T)$:

$$\rho(T) = f(T)\rho_D(T), \tag{S11}$$

with the transmission distribution for a diffusive conductor

$$\rho_D(T) = \frac{G_D}{2G_Q T \sqrt{1-T}}, \tag{S12}$$

where $f(T)$ is implicit:

$$T = \frac{1}{\cosh^2\left(\frac{\mu}{2}\right)}; \quad \mu = \operatorname{arccosh} Y - \frac{G_T}{G_D}\sqrt{Y^2 - 1}\cos(\pi f); \quad Y = \frac{G_D \pi f}{G_T \sin(\pi f)}. \tag{S13}$$

If $G_D \gg G_T$, we expect all $T$ to be very small, i.e. the distribution must turn to zero above some critical value $T_c$ with $T_c \ll 1$. In this case,

$$\mu = \ln\frac{4}{T} = \operatorname{arccosh} Y - \frac{G_T}{G_D}\sqrt{Y^2 - 1}\cos(\pi f). \tag{S14}$$

For $T \ll 1$, we have $Y \gg 1$ and then $\sqrt{Y^2 - 1} \approx Y$ and $\operatorname{arccosh} Y \approx \ln 2Y$. We get thus

$$\ln\frac{4}{T} = \ln 2Y - \frac{G_T}{G_D} Y \cos(\pi f). \tag{S15}$$

Then we can combine equation S13 and S17, take $G_D \gg G_T$ into account and get

$$\frac{G_T}{TG_D} = \frac{\pi f}{\sin \pi f} e^{-\pi f \cot \pi f}. \tag{S16}$$

The maximum value that the right-hand side can reach is 1/e (achieved at $f = 0$), and thus the maximum possible transmission eigenvalue is $T_C = \frac{eG_T}{G_D} \ll 1$. We see indeed that all possible transmission eigenvalues are small, and the distribution is given by the combination of equations (S11), (S12), and (S16).

For our previous analysis, using this language, we use the distribution $\rho(T) = \delta(T - T')$, where $T'$ is determined by the total transmission and the number of transport channels. The distribution (S16) is different, in particular, it even diverges at $T = 0$. However, since all possible transmission eigenvalues are small, we have $\langle T(1 - T) \rangle \approx \langle T \rangle$, and the average value of the shot noise is determined by the average transmission eigenvalue, which we determine from the experiments. Thus, as soon as all transmission eigenvalues are small, the presence of the distribution does not affect our analysis.

### XI: Sample information

In this work, we measured noise on three Nb/Al-AlO$_x$/Nb SIS samples (sample 1, sample 2 and sample C1) and one NbN/oxide/Ag SIN junction (sample N1), see table S1 and table 1. Sample 1 and sample 2 have a 0.82 μm$^2$ junction area. They are fabricated on a fused silica wafer. First, a 120 nm thick niobium layer is deposited by a DC sputtering process. On top of this niobium layer, a 8 nm thin Al layer is deposited in situ, which is subsequently oxidized using a dynamic oxidation at a fixed chamber pressure of 1.6 Pa. The AlO$_x$ barrier thickness, and thus the normal-state resistance, is tuned with the oxidation time. On top of this oxide, again a niobium layer of 160 nm is deposited in situ, forming together the SIS trilayer. The trilayer is subsequently patterned using a mix-and-match optical and electronic lithography process, followed by a dry plasma etch of the top niobium layer in a fluorine based plasma, thus defining the micron-sized SIS junction. A self-aligned ~ 200 nm thick SiO$_2$ layer is deposited on the same resist mask for electrical isolation, after which a third niobium layer is defined on top of the SiO$_2$ layer, permitting electrical contact to the top electrode of the SIS junction. In a final step, gold contact pads are defined facilitating wire bonding of the devices.
Sample C1 has 25 μm$^2$ junction area and is bought from STAR Cryoelectronics.

The NbN/oxide/Ag junction sample N1 has a 300×300 μm$^2$ junction area. The NbN film is synthesized by sputtering an Nb target in an Ar/N2 gas mixture using reactive DC magnetron sputtering on a (100)-oriented single-crystalline MgO substrate. The junction is fabricated by first depositing a 300 μm wide NbN strip at 600℃. Then, the NbN is oxidized in air at 250 ℃ for 1hour 50min. Finally, 300 μm wide cross strips of Ag are deposited at room temperature, and these cross sections form our SIN junctions. Two junctions are fabricated on one substrate at the same time.

We also summarize all our experimental results at different temperatures, different magnetic fields and different samples in table S1. All of our experimental results indicate that $q = 1e$ noise inside the gap.

### XII: Results in previous work

According to our theory, the shot noise will decrease with decreasing $\tau$. To compare our theory to results from literature ([12-15,17,34,35] for main text), we first estimate $\tau$ from given parameters, and compare the predictions of our model with results. We find that $q > 1e$ was always reported in systems with large $\tau$ and $q = 1e$ was only reported in systems with small $\tau$, in agreement with our model – with one outlier, as discussed below. We summarize the results and estimated $\tau$ in previous work and our work in the Table S2. The effective charge is estimated by $\tau = (G_0 R_j)^{-1}$ for single channel systems [12-14,34,35] and by the same method we discussed in the main text for multiply channel systems [15,17].

We can also plot the numbers from the table in Fig. S9. We find that $q > 1e$ is reported in systems with large $\tau$ (roughly $\tau > 10^{-4}$). And when $\tau$ is small, $q = 1e$ is usually measured. These results fit our theory well. The only exception is Ref. [17]: $q > 1e$ was reported with very small $\tau \sim 10^{-8}$. This result is from cuprates, where things tend to be more complicated. Possible scenarios are: (i) preformed pairs that need a completely different theory, (ii) noise from noise centers in the charge transfer layers, or (iii) special tunneling channels though impurities/dopants.

**XIII: Uncertainty of experimental results**

In our experiment, we measured voltage noise power $S_V$, temperature $T$, differential resistance $R_{diff}$. And we use these to calculate effective charge $q$, and effective HEMT power gain $Gain_{HEMT}$. The uncertainty of these are shown in table S3.

The uncertainty of $S_V$ is estimated from the noise measurement at similar bias outside the superconducting gap. The temperature is measured by a calibrated thermometer near the sample and exchange gas is used to ensure that the temperature is homogenous. The uncertainty of $T$ is estimated from the temperature vs time during the measurement. The uncertainty of $R_{diff}$ is estimated from the fluctuation of voltage measured by a lock-in. And the uncertainty of $Gain_{HEMT}$ is obtained from the standard deviation of the calibration fitting line.

The uncertainty of $q$ is calculated from the uncertainty of $S_V$ and equation: $S_V = 2q|I|\coth(q|V|/2k_B T)R_{diff}^2$. The uncertainty of the effective charge is mainly from the noise power $S_V$ measurement instead of resistance measurement. $S_V$ will decrease with decreasing voltage but our uncertainty for noise, $\Delta S_V$, is nearly constant for all the bias due to the noise background from circuit. Thus, the uncertainty of the effective charge $\Delta q \sim \frac{\Delta S_V}{I}$ is larger at small bias current than large bias.

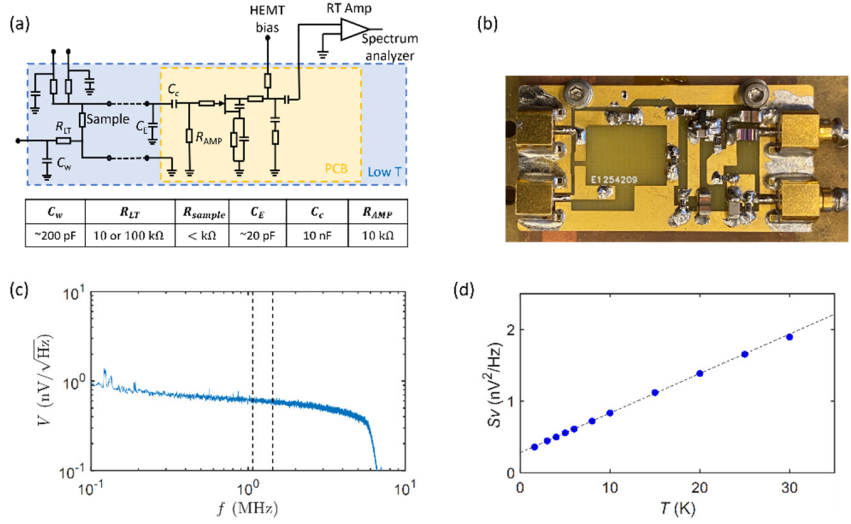

Fig. S1: Amplifier calibration. (a) Schematic of amplifier. (b) Photograph of amplifier. (c) Typical noise spectrum density for thermal noise of a 1 kΩ resistor at 1.59 K. Dashed lines demarcate the frequency range used for our data analysis. (d) Thermal noise of the resistor as a function of temperature. Blue dots are experimental data and black dashed line is the calibration fitting line.

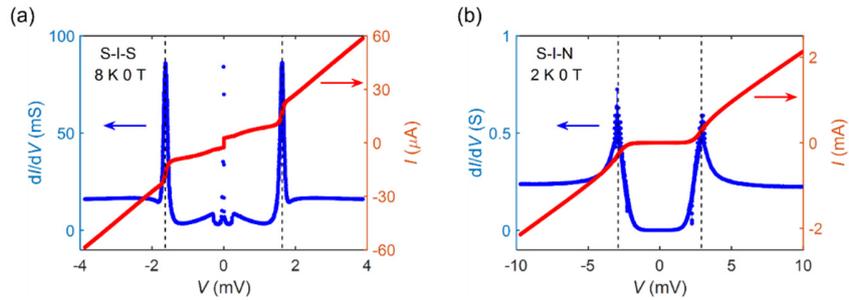

Fig. S2: Nonlinear $I$-$V$ (red lines) and $dI/dV$ (blue lines) curves of (a) a SIS junction at 8 K (sample C1) and (b) a SIN junction (sample N1) at 2 K. The measurements are done before noise measurements. Black dashed lines indicate the superconducting gap. The nonlinear $I$-$V$ shape leads to a double peak structure of the noise power ($S_V$) inside the gap.

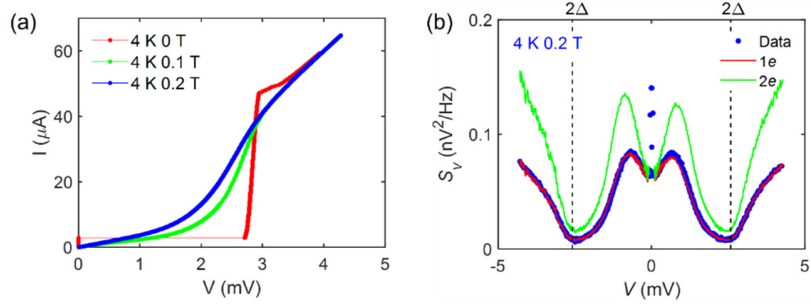

Fig. S3: Impact of magnetic field on *I-V* and $S_V$-*V* curves for an SIS junction, sample C1. (a) *I-V* curve at 4 K with magnetic field $B = 0$ (red), 0.1 T (green), and 0.2 T (blue). The supercurrent disappears when the magnetic field is applied to sample. (b) $S_V$-*V* curve at 4 K and 0.2 T. Blue dots are experimental data, and the red and green lines are the 1*e* and 2*e* theory lines, respectively. Dashed lines indicate the superconducting gap. Noise at 4 K and 0.2 T still shows *q* = 1*e*.

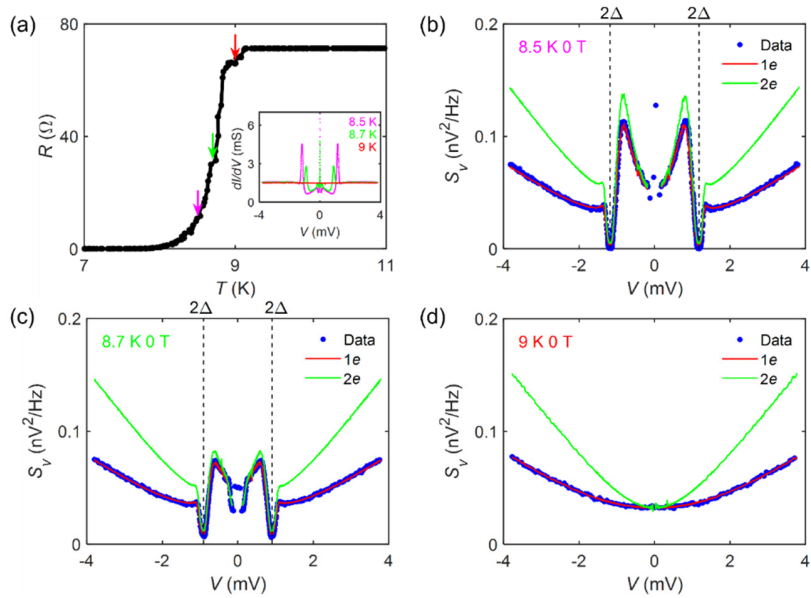

Fig. S4: Characteristics of a SIS junction sample C1 near $T_C$. (a) *R-T* curve. Arrows indicate the temperatures at which noise was measured (8.5 K, 8.7 K, and 9 K). Inset: d*I*/d*V* curves for these temperatures. Noise results at 8.5 K, 8.7 K, and 9 K are shown in (b), (c), and (d). Blue dots are experimental data, and the red and green lines are the 1*e* and 2*e* theory lines, respectively.

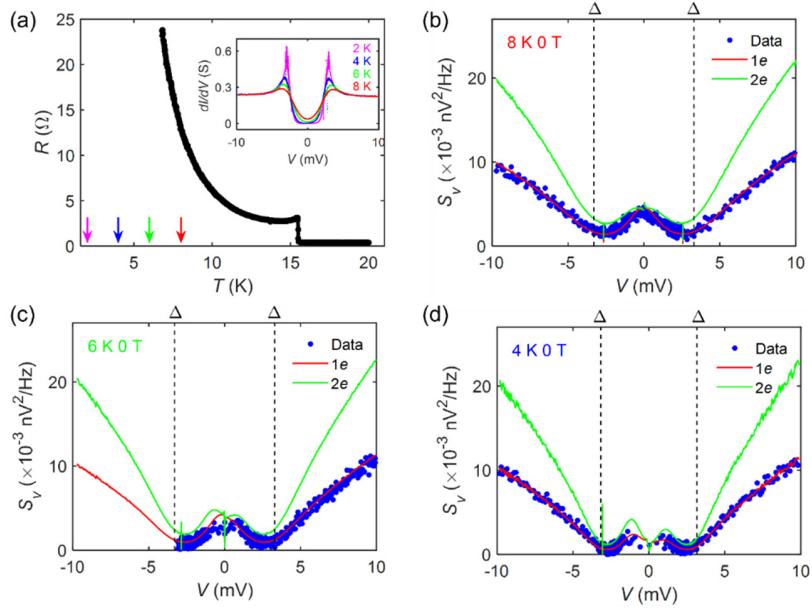

Fig. S5: Characteristics of a SIN junction, sample N1, below $T_C$. (a) $R$-$T$ curve. Arrows indicate the temperatures at which noise was measured (2 K, 4 K, 6 K, and 8 K). Inset: tunneling spectroscopy at these temperatures. Noise results at 8 K, 6 K, and 4 K are shown in (b), (c), and (d). Blue dots are experimental data, and the red and green lines are the 1$e$ and 2$e$ theory lines, respectively. Results of 2 K are shown in Fig. 4 of the main text. All the results indicate $q = 1e$.

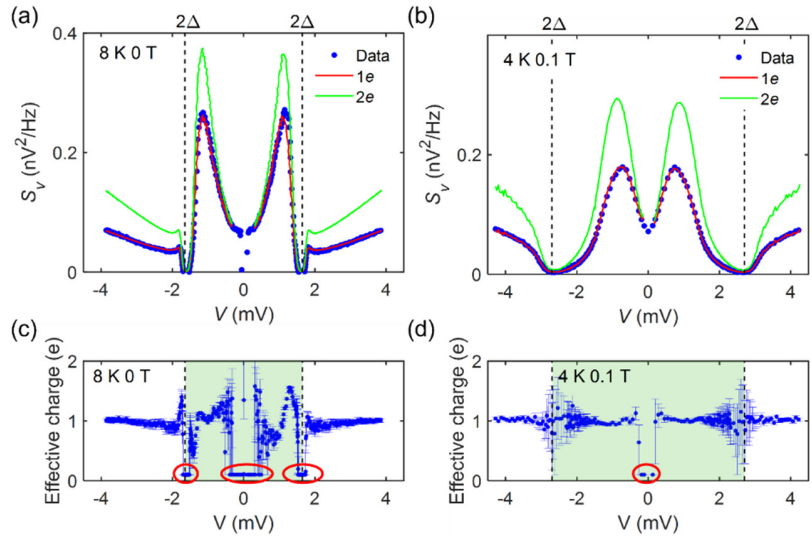

Fig. S6: Noise and effective charge. (a,b) Noise power at (a) 8 K and 0 T, and (b) 4 K and 0.1 T for sample C1. Blue dots are experimental data, and the red and green lines are the 1$e$ and 2$e$ theory lines, respectively. (c,d) Effective charge vs voltage at (c) 8 K and 0 T, and (d) 4 K and 0.1 T. Effective charges are calculated using Eq. (9). Green areas indicate the data used to calculate the averaged effective charge and red circles indicate the data that were excluded from the calculation.

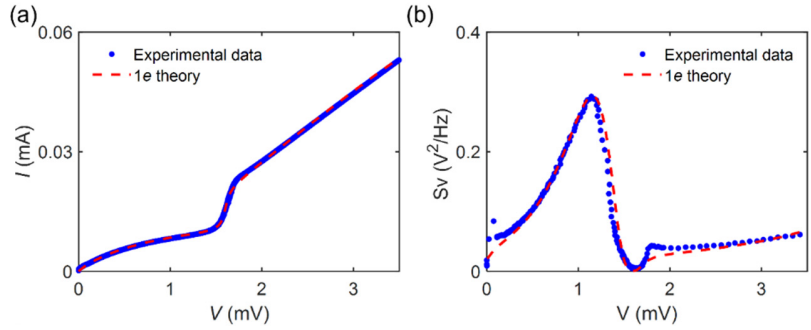

Fig. S7: Normal tunneling in a SIS junction. (a) *I-V* curve and (b) noise power of a SIS junction, sample C1, at 8 K and 0 T. Blue dots are experimental data and red dotted line is calculated from the Bardeen tunneling theory for normal tunneling.

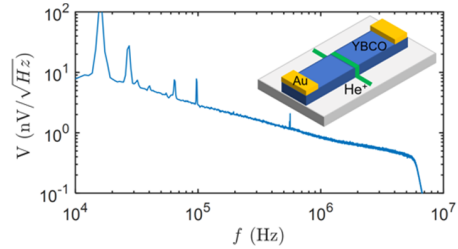

Fig. S8: Shot noise in a Josephson junction made of the high-$T_c$ superconductor $YBa_2Cu_3O_7$ (YBCO) fabricated using a focused ion ($He^+$) beam. The spectrum is dominated by $1/f$ noise and accurate shot noise cannot be calculated. Inset: schematic of the sample structure.

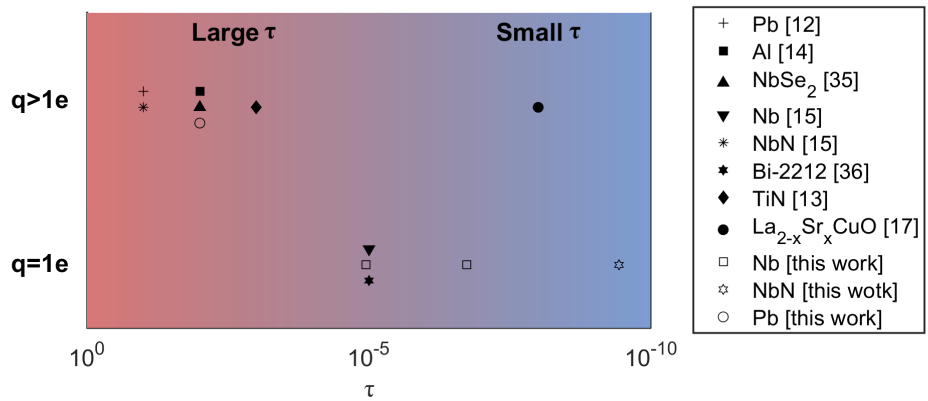

Fig. S9: Reported shot noise and estimated transparency in previous work. Reference numbers are related to the references in the main text.

**Table S1 Summary for noise measurement below $T_c$**

| Sample structure | S-I-S Nb/AlOx/Nb | | | | | | | S-I-N NbN/oxide/Ag |
|---|---|---|---|---|---|---|---|---|
| Sample name | Sample 1 | | Sample 2 | | Sample C1 | | | Sample N1 |
| Zero resistance $T_c$(K) | ~ 8.2 K | | | | ~ 8 K | | | ~ 15 K |
| $T_{measured}$ (K) | 7 | 8.1, 7.8 | 7 | 8.1 | 4 | 8 | 8.5, 8.7, 9 | 2, 4, 6, 8 |
| B (T) | 0.1, 0.25, 0.37 | 0 | 0.1, 0.2, 0.25, 0.3, 0.37, | 0 | 0.1, 0.2 | 0, 0.1 | 0 | 0 |
| Effective charge | ~ 1e | | | | | | | |

**Table S2 Estimated transparency in previous work**

| $q > 1e$ | | $q = 1e$ | |
|---|---|---|---|
| $\tau$ | Materia | $\tau$ | Material |
| $\sim 10^{-1}$ | Pb[12] | $\sim 10^{-5}$ | Nb[15] |
| $\sim 10^{-1}$ | Al[14] | $\sim 10^{-5}$ | B-2212 [35] |
| $\sim 10^{-2}$ | NbSe$_2$[34] | $\sim 10^{-5}$ | Nb [this work] |
| $\sim 10^{-1}$ | NbN[15] | $\sim 10^{-7}$ | Nb [this work] |
| $\sim 10^{-3}$ | TiN[13] | $\sim 10^{-10}$ | NbN[this work |
| $\sim 10^{-8}$ | La$_{2-x}$Sr$_x$CuO [17] | | |

* reference numbers are for main text

**Table S3 Uncertainty of experimental results**

| $S_V$ | $T$ | $R_{diff}$ | $q$ | $Gain_{HEMT}$ |
|---|---|---|---|---|
| 0.002 nV$^2$/Hz | 0.01 K | ~1% | Shown in figure | 0.01 |

[1] Y. M. Blanter and M. Büttiker, *Shot Noise in Mesoscopic Conductors*, Phys Rep **336**, 1 (2000).

[2] G. E. Blonder, M. Tinkham, and T. M. Klapwijk, *Transition from Metallic to Tunneling Regimes in Superconducting Microconstrictions: Excess Current, Charge Imbalance, and Supercurrent Conversion*, Phys Rev B **25**, 4515 (1982).

[3] Y. Ronen, Y. Cohen, J. H. Kang, A. Haim, M. T. Rieder, M. Heiblum, D. Mahalu, and H. Shtrikman, *Charge of a Quasiparticle in a Superconductor*, Proceedings of the National Academy of Sciences **113**, 1743 (2016).

[4] K. M. Bastiaans, T. Benschop, D. Chatzopoulos, D. Cho, Q. Dong, Y. Jin, and M. P. Allan, *Amplifier for Scanning Tunneling Microscopy at MHz Frequencies*, Review of Scientific Instruments **89**, 93709 (2018).

[5] F. Massee, Q. Dong, A. Cavanna, Y. Jin, and M. Aprili, *Atomic Scale Shot-Noise Using Cryogenic MHz Circuitry*, Review of Scientific Instruments **89**, (2018).

[6] R. C. Dynes, V. Narayanamurti, and J. P. Garno, *Direct Measurement of Quasiparticle-Lifetime Broadening in a Strong-Coupled Superconductor*, Phys Rev Lett **41**, 1509 (1978).

[7] B. Müller, M. Karrer, F. Limberger, M. Becker, B. Schröppel, C. J. Burkhardt, R. Kleiner, E. Goldobin, and D. Koelle, *Josephson Junctions and SQUIDs Created by Focused Helium-Ion-Beam Irradiation of YBa2Cu3O7*, Phys Rev Appl **11**, 044082 (2019).

[8] Y. V. Nazarov and Y. M. Blanter, *Quantum Transport: Introduction to Nanoscience*, Cambridge University Press (2009).